# ANOMALOUS ELECTRONIC SUSCEPTIBILITY IN $Bi_2Sr_2CuO_{6+\delta}$

Guénaëlle LE BRAS-JASMIN[1], Zorica KONSTANTINOVIC[1], Jean Pierre CARTON[1], Dorothée COLSON[1], Anne FORGET, Frédéric JEAN[2,3], Gaston COLLIN[3], Yves DUMONT[4] and Claude AYACHE[1]

[1]Service de Physique de l'Etat Condensé, DRECAM / DSM, CEA Saclay, 91191 Gif sur Yvette, France
[2]LEMHE, CNRS UMR 8647, Bât 415, Université Paris Sud, 91405 Orsay, France
[3]LLB, DRECAM / DSM, CEA-CNRS, CEA Saclay ; 91191 Gif sur Yvette, France
[4]LMOV, Université de Versailles-Saint Quentin en Yvelines, 45 Av. des Etats Unis, 78035 Versailles, France

We report magnetic susceptibility performed on overdoped $Bi_2Sr_2CuO_{6+\delta}$ powders as a function of oxygen doping $\delta$ and temperature T. The decrease of the spin susceptibility $\chi_s$ with increasing T is confirmed. At sufficient high temperature, $\chi_s$ presents an unusual linear temperature dependence $\chi_s = \chi_{s0} - \chi_1 T$. Moreover, a linear correlation between $\chi_1$ and $\chi_{s0}$ for increasing hole concentration has been displayed. These non conventional metal features will be discussed in terms of a singular narrow-band structures.

INTRODUCTION

It is now believed that the unusual normal state properties in the cuprates reflect the electronic structure that underlies high $T_c$ superconductivity. During several years, the overdoped regime, considered as dispalying the "classical" electronic properties of a normal metal, has been less studied. Recent theoretical and experimental results suggest a departure from a conventional FL not only for underdoped range but also in the overdoped one[1,2,3]. A strong increase in the electronic density of states (DOS) with doping has been proposed as a possible origin for these unusual properties in the overdoped regime[4]. Such an hypothesis can be fruitfully investigated by comparison with recent progress obtained through high resolution ARPES measurement in this overdoped range.

In this paper, we report measurements of bulk magnetic susceptibility as a function of temperature and doping in the normal state of overdoped $Bi_2Sr_2CuO_{6+\delta}$ (Bi-2201) powders.

EXPERIMENTAL RESULTS AND DISCUSSION

The compound Bi-2201 is a convenient system for studying the normal state in the overdoped regime. Indeed, in its as grown form (ie non susbstitued), this compound lies naturally in the overdoped range. It also belongs a low $T_{cmax}$(20K), allowing us to study the normal state properties over an extended temperature range. Polycrystalline samples were prepared by the classical solid state reaction method[5]. The absolute content of oxygen nonstoichiometry $\delta$ of the compound $Bi_2Sr_2CuO_{6+\delta}$ have been investigated by thermogravimetric techniques from 300°C to 700°C with oxygen partial pressure $P(O_2)$ ranging from $10^{-4}$ atm to 1 atm under thermodynamic equilibrium conditions[6]. We have explored the T-$\delta$ phase diagram for oxygen concentration varying from $\delta$=0.09 to 0.18 $\pm$0.05. The hole concentration p in the $CuO_2$ planes has been estimated by thermoelectric power measurements performed on sintered samples made from the same powder and submitted to the same oxygen treatment[7]. We found that p changes from 0.20 to 0.29 causing a decrease from $T_c \cong$16K to $T_c$<1.5K.

Figure 1 shows the temperature dependence between 50K and 300K of the normal-state susceptibility under 5T for the various hole concentrations. $\chi$(T) close to optimal doping sample (p ~0.20), is moderately temperature independent with a positive slope around T$\cong$100K which should may be a reminiscence of the pseudogap effect. On increasing hole concentration, $\chi$ (T) increases with decreasing temperature. The negative slope $d\chi/dT$ observed at room temperature (RT) is a general trend in all overdoped cuprates.

The spin susceptibility is obtained by susbstracting the Van Vleck paramagnetism and the core diamagnetism essentially temperature independent (the values of these contributions are listed in the table 1 of the ref.2) and can be fitted by the following expression $\chi_s(T) = \chi_{s0} - \chi_1 T$.

The linear temperature dependence $-\chi_1 T$ is unusual. It does not correspond to the first $T^2$ temperature correction observed in a conventional metal. It seems that it exists in the overdoped cuprates a general behavior of the normal states properties with a linear T dependence[7,8,9,10].

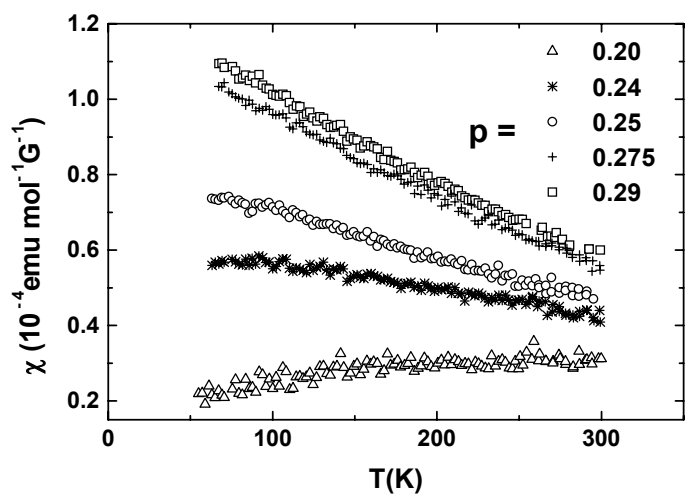

FIGURE 1

Magnetic susceptibilities of the polycrystalline samples $Bi_2Sr_2CuO_{6+\delta}$ for various hole doping p under H=5T.

In fig. 2(a), we have reported $\chi_{s0}$ (left scale) and $\chi_1$(right scale) as a function of p, in the overdoped regime. Remarkably, $\chi_{s0}$ as well as $\chi_1$ have a similar behavior as a function of p. At p~0.20, corresponding to the less overdoped Bi-2201 sample, $\chi_1$ is nearly 0 (as shown above in fig. 1, $\chi$ is nearly temperature independent near the optimal doping). With increasing hole doping, both $\chi_{s0}$ and $\chi_1$ drastically increase.

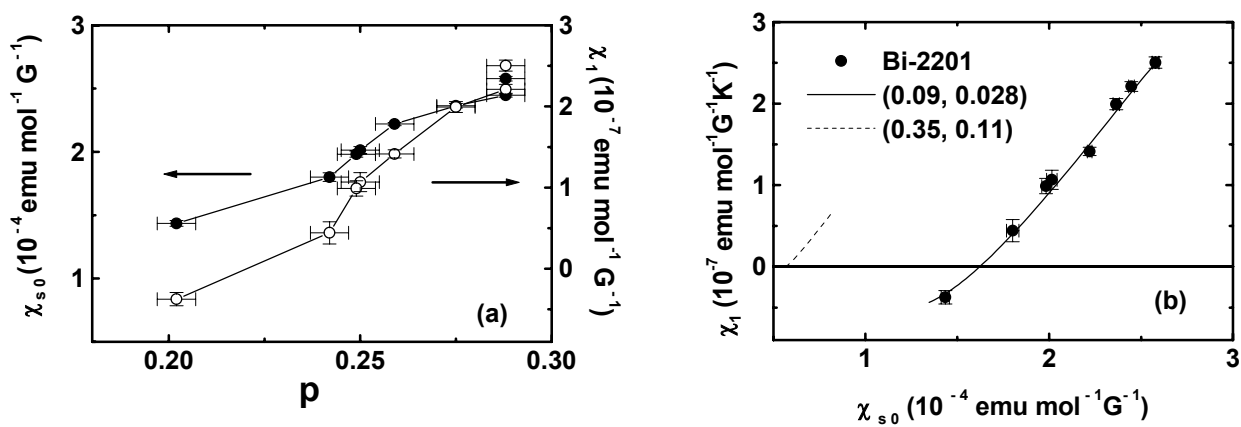

FIGURE 2

(a) hole doping dependence of the two contributions to the electronic susceptibility $\chi_{s0}$ (solid circle) and $\chi_1$ (open circle); (b) the susceptibility calculated in the frame of tight-binding one band model with the nearest-neighbor and next-nearest-neighbor interactions t=0.35 eV, t'=0.11eV (dashed line) and t=0.09 eV, t'= 0.028 eV (solid line).

In figure 2b, we have reported $\chi_1$ versus $\chi_{s0}$ for Bi-2201 (solid circle), for different oxygen content. Remarkably, for all hole doping a linear increase of $\chi_1$ with $\chi_{s0}$ (for increasing hole concentration) is observed. This linear correlation between $\chi_1$ and $\chi_{s0}$ on a large range of hole doping seems to be a universal trend in the overdoped cuprates[11]. It means that the hole doping dependence of $\chi_1$, shown in figure 2a, arises in fact from the hole doping dependence of $\chi_{s0}$. Thus the growing of $\chi_s$ as a function of hole doping (see fig.2a) should reflect the presence of a peak in the DOS. Such a deduction is in good agreement with ARPES measurements on HTSC that have allowed to identify the presence of saddle points centered at k= ($\pi$, 0) and (0, $\pi$) in the band structure which give rise to van Hove singularities (VHS) in the density of states (DOS) near the Fermi level. In the case of overdoped Bi-2201 this VHS lies below $E_F$ for all doping levels[12,13,14], which means that the Fermi Surface (FS) retains its hole like character centered at ($\pi,\pi$). We have therefore used a model compatible with the Van Hove scenario and ARPES measurements. By this way, we have reproduced the doping dependence of the spin susceptibility in terms of a tight-binding one band model of the $CuO_2$ square lattice involving VHS, expressed by

$$E(k) = -2t(\cos k_x a + \cos k_y a) + 4t' \cos k_x a \cos k_y a$$

where a is the lattice parameter, t and t' are respectively the nearest-neighbor and next-nearest neighbor interactions and $E_{VHS} = -4t'$. For t'=0, the saddle point is at the band center, the Fermi level ($E_F$) coincides with it for the half-filled case. It means that the change of topology of FS from hole-like to electron like appears in this case at zero doping. As t' increases, the VHS is pushed close to the bottom of the band and the doping required to keep the saddle point at $E_F$ also increases. The hole doping p is directly related to the distance $D=E_F-E_{vhs}$ and reported for Bi-2201 in ref. 15. It has also been shown in this ref. that the change of topology of FS from hole-like-to electron like may arise in Bi-2201 for p ~ 0.29 corresponding to a ratio t/t' ~ 3.2.

The spin susceptibility is obtained by using the following expression

$$\chi_s = \mu_B^2 \int n(E)[-\partial f(E)/\partial E]dE$$

where n(E) is the density of states and $f(E)$ is the fermi function.

As $\chi_s$ is well described for all hole doping by a T-linear law, we have therefore calculated $\chi_s$ (100K) and $\chi_s$(300K) for different values of p. Simulations have been made for 2 sets of values (t, t') corresponding to a ratio t/t' ~3.2. The values of $\chi_{s0}$ and $\chi_1$ obtained by this way are shown in fig 2b. The slope $d\chi_1/\chi_{s0}$ is well reproduced for both set of values (t, t'). However the spin susceptibility calculated with t =0.35 eV and t' =0.11 eV (dashed line) supported by ARPES measurements, is largely shifted from the experimental data (solid circle). Smaller values of t and t' (0.09 eV, 0.28 eV) are required to well reproduce the variation of $\chi_1$ *vs* $\chi_{s0}$ in Bi-2201 (solid line).

CONCLUSION

In conclusion, the present study features that in the overdoped regime, the magnetic susceptibility as a function of doping p and temperature T shows a departure from the conventional FL which can be interpreted by the presence of a van Hove singularity near the Fermi level in the DOS, in good agreement with the ARPES observations.

ACKNOWLEDGMENTS

The authors thank J. Bok, J. Bouvier, M. Norman, C. Pépin and M. Roger for fruitful discussions, and L. Le Pape for his technical support.